\documentstyle[aps,prl,multicol]{revtex}
\input{epsf}
\begin{document}
\draft
\preprint{\today}
\title{From the Fermi glass towards the Mott insulator in one dimension:\\ 
Delocalization and strongly enhanced persistent currents}
\author{Peter Schmitteckert$^{1,2}$, Rodolfo~A. Jalabert$^{1}$, Dietmar 
Weinmann$^{3}$ and Jean-Louis Pichard$^{2}$}
\address{1:~Institut de Physique et Chimie des Mat\'eriaux de Strasbourg, 
            23 rue du Loess, 67037 Strasbourg cedex, France\\
         2:~CEA, Service de Physique de l'Etat Condens\'e,
           Centre d'Etudes de Saclay, F-91191 Gif-sur-Yvette, France\\
         3:~Institut f\"ur Physik, Universit\"at Augsburg, 86135 
          Augsburg, Germany}
\maketitle
\begin{abstract}
When a system of spinless fermions in a disordered mesoscopic ring becomes 
instable between the inhomogeneous configuration driven by the random 
potential (Anderson insulator) and the homogeneous one driven by 
repulsive interactions (Mott insulator), the persistent current can be 
enhanced by orders of magnitude. This is illustrated by a study 
of the change of the ground state energy under twisted boundary 
conditions using the density matrix renormalization group algorithm. 
\end{abstract}
\begin{multicols}{2}
\narrowtext
  For a fixed kinetic energy ($t\!=\!1$), the interplay between the disorder 
($W$) and a repulsive interaction ($U$) is one of the central 
problems of mesoscopic physics~\cite{Moriond}. For instance, its 
understanding appears as a necessary step towards explaining why the 
measured persistent currents~\cite{Levy} $J$ in mesoscopic rings are 
about two orders of magnitude larger than the value predicted by the 
noninteracting theory~\cite{nonint}. This discrepancy has 
prompted to study the role of the interaction using perturbation 
theory~\cite{eckern}, assuming a Luttinger liquid and using 
renormalization group~\cite{gs} for the disorder, or considering 
the limit where the rotational invariance of the interacting system 
is broken by the disorder~\cite{AetW}. On the other hand, exact numerical 
calculations on small $1d$ models of spinless fermions have led to the 
conclusion that repulsive interactions cannot significantly~\cite{AetB} 
enhance $J$, being on the contrary rather detrimental~\cite{BPM} to the 
expected effect. It was then argued that this is an artifact of spinless $1d$ 
models, and that the full problem with spin does exhibit~\cite{gs} 
the expected enhancement. We show in this Letter that even without spin, 
repulsive interactions can give rise to a large enhancement associated 
with a charge reorganization of the ground state. The interaction for 
which the ground state is reorganized fluctuates from sample to sample. 
Thus, the effect could not be seen in the previous studies of 
the ensemble averages, but it becomes very striking if one studies the 
orbital response of individual mesoscopic samples as a function of $U$.

 Another motivation to re-examine the ground state of one 
dimensional spinless fermions stems from Shepelyansky's 
proposal concerning the role of interactions in an Anderson 
insulator \cite{shepelyansky}: two interacting particles (TIP) 
can be localized on a length $L_2$ much larger than the one-particle 
localization length $L_1$. Large ratios $L_2/L_1$ characterize the 
states in the bulk of the TIP-spectrum. However, a TIP ground state built 
from two one-particle states localized far away from each 
other cannot be reorganized by a short range interaction. 
Interesting effects on the ground state properties require 
reasonably high filling factors. For the finite density system, 
a Fermi liquid approach was suggested by Imry~\cite{Imry} and 
later developed in Refs. \cite{oppen,js}, reducing the problem 
to the study of the delocalization of a few quasi-particles above 
the Fermi sea. From a scaling argument, one obtains that two  
quasi-particles need an excitation energy $\epsilon 
\approx t$ (irrelevant for the usual applications in solid state 
physics) in order to recover the enhancement factor $L_2/L_1$ of 
two (excited) bare particles. This quasi-particle approach may be 
appropriate if there is no dramatic reorganization of the ground 
state by the interaction. We want to consider another limit: 
when the Fermi liquid becomes instable as the Mott insulator 
is approached. To avoid any confusion, let us say that the Fermi liquid 
will be better named a Fermi glass in the disordered limit which we 
consider, characterized by Anderson localization without interaction.
By Mott insulator, we mean a strongly correlated array of 
charges which is pinned and distorted by the random lattice.  
 
  For the TIP problem (symmetric states of two electrons with opposite 
spins and on-site interaction), the ratio $L_2/L_1$ is maximum~\cite{wwp} 
when the system becomes instable between two limits: free bosons for $U\!=0\!$ 
and hard core bosons for $U\!=\!\infty$. When $U\!\approx\!t$, the mixing of 
the one-particle states by the interaction is maximum and the 
system becomes weakly chaotic with critical spectral 
statistics for system sizes $M\!\approx\!L_1$. This critical regime 
cannot be described by perturbation expansions in powers of $U$ or 
$t^2/U$, starting from the free boson and hard-core boson limits 
respectively. Therefore, approaches starting from simple 
limits for the ground state (Luttinger liquid for weak 
disorder~\cite{gs2}, Mott insulator for strong interaction, Fermi 
glass for weak interaction and large disorder) may not be valid for 
$U\!\approx\!t$ and large disorder, say $k_{\rm f}l\!\approx\!1$ 
($k_{\rm f}^{-1}$ characterizes the spacing between the electrons in the 
clean system and in $1d$ the elastic mean free path is $l\!=\!L_1$).
 
   This suggests us to study the many body ground state in the most 
chaotic case, between the limits which correspond to the free 
and hard-core particles in the archetypical TIP example. If $M$ 
denotes the number of sites of a $1d$ lattice with $N$ spinless fermions, 
we expect a maximum delocalization when the parameters ($U, t, W, N$ and 
$M$) are chosen in such a way that the system becomes instable between 
two insulating limits with different charge configurations: Anderson 
insulator in the Fermi limit and Mott insulator in the strongly correlated 
limit. The Anderson insulator yields an inhomogeneous charge configuration: 
strongly localized states are populated with the restriction imposed by 
Pauli principle. The Mott insulator yields a more or less 
homogeneous charge configuration. The crossover between those two 
limits will yield a profound spatial reorganization of the ground 
state, making the system more sensitive to any external perturbation 
as a twist in the boundary conditions or a flux $\Phi$ in a ring.  
 
%
%

 To illustrate this, we consider spinless fermions on a disordered
chain with nearest neighbor (NN) interaction
$$
H=-t \sum_{i=1}^{M} (c_i^{\dagger} c_{i-1} + c^{\dagger}_{i-1}c_i) 
+\sum_{i=1}^{M} v_i n_i + U \sum_{i=1}^{M} n_i n_{i-1}
$$
and twisted boundary conditions, $c_0=\exp(i\Phi) c_M$. The operators $c_i$ 
($c^{\dagger}_{i}$) destroy (create) a particle on site $i$ and 
$n_i=c^{\dagger}_ic_i$ is the occupation operator. The on-site energies
$v_i$ are drawn from a box distribution of width $W$. Setting $t\!=\!1$, the 
strength of the disorder $W$ and the interaction $U$ are measured in 
units of the kinetic energy scale. 
Though strong long-range interaction always imposes an homogeneous 
configuration, we restrict the study to interaction between NN. 
This is why we begin to discuss half filling, where 
the ground state will be a periodic array of charges located 
on the even or odd sites of the chain when $U \rightarrow \infty$. 
This is the homogeneous configuration driven by the interaction. On the 
contrary, when $U \rightarrow 0$, an inhomogeneous configuration is obtained
for a sufficiently strong disorder.
 
 The numerical results are obtained with the density matrix renormalization
group (DMRG) algorithm \cite{dmrg}. 
The calculation of ground state properties of interacting 
fermions in disordered one-dimensional systems is 
\begin{figure}[tbh]
\centerline{\epsfxsize=3in\epsffile{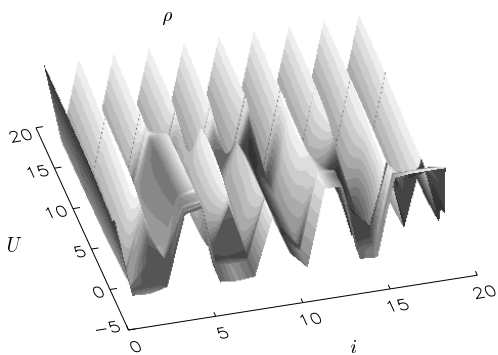}}
\vspace{2mm}
\caption[fig1]{\label{density} Charge configuration for a typical sample 
(d of Fig.~\ref{D(U)}) for $N=10$ particles on $M=20$ sites at $W=9$.}
\end{figure}

\begin{figure}[tbh]
\centerline{\epsfxsize=3in\epsffile{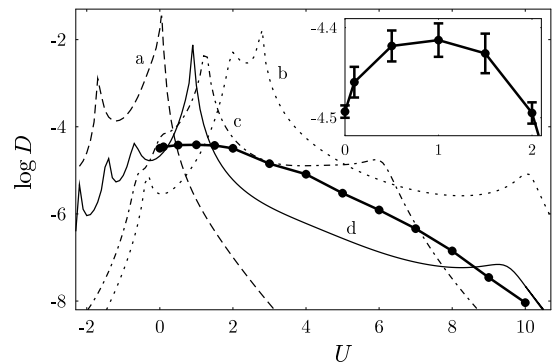}}
\vspace{3mm}
\caption[fig2]{\label{D(U)} Phase sensitivity $D(U)$ for four different 
samples with $N=10$, $M=20$ and $W=9$ in (decimal) logarithmic scale.
$W$ and $U$
are measured in units of $t$. Thick dots and inset: average of 
$\log(D)$.}
\end{figure}
\noindent
possible with an accuracy comparable to exact diagonalization, but
for much larger systems (where we keep up to 2000 states per block 
\cite{peter}).

 The reorganization of the ground state induced by the NN
repulsion is shown in Fig.~\ref{density}, where the density $\rho$ 
(expectation value of $n_i$) is plotted as a function of $U$ and site 
index $i$. To favor the inhomogeneous configuration, the disorder is taken 
large ($W\!=\!9$) and $L_1\!\approx\!100/W^2$ is of order of the mean 
spacing $k_{\rm f}^{-1}\!=\!2$ between the charges. For $U\!\approx\!0$, 
one can see 
a strongly inhomogeneous and sample dependent density, while for large $U$ 
a periodic array of charges sets in. 
These two limits are separated by a sample dependent crossover regime. 
For certain random configurations, the periodic array is quickly obtained by 
a weak repulsive interaction, while one needs a strong interaction for other 
samples. 

 To measure the delocalization effect associated to this change of 
configuration, we study the phase sensitivity of the ground state. 
The energy difference between periodic ($\Phi\!=\!0$) and anti-periodic 
($\Phi\!=\!\pi$) boundary conditions, $\Delta E = (-)^N (E(0)\!-\!E(\pi))$ 
conveys similar information, in the localized regime, as other measures 
of the response of the ground state to an infinitesimal flux threading the 
ring: the Kohn curvature (charge stiffness) $\propto E^{''}(\Phi\!=\!0)$ 
and the persistent current $J \propto -E^{'} (\Phi\!=\!0)$. 
For strictly $1d$ systems, the sign of $E(0)\!-\!E(\pi)$ simply depends 
on the parity of $N$, and the factor $(-)^N$ makes $\Delta E$ positive. 

 The phase sensitivity, defined by $ D(U)=(M/2) \Delta E$, is shown in 
Fig.~\ref{D(U)} for four samples at half filling with $W\!=\!9$. Both 
for $U\!\approx\!0$ and $U\!\gg\!1$, $D(U)$ is exponentially 
small, but sharp peaks appear at sample dependent values $U_c$, 
where $D(U_c)$ in certain samples can be 4 {\em orders of magnitude 
larger than for free fermions}. Remarkably, the curves for each 
sample do not present any singularity at $U\!=\!0$ which could allow 
to locate the free fermion case. Peaks can be seen at different sample 
dependent values of $U$ (positive or negative). Attractive interactions favor 
 
\begin{figure}[tbh]
\centerline{\epsfxsize=3in\epsffile{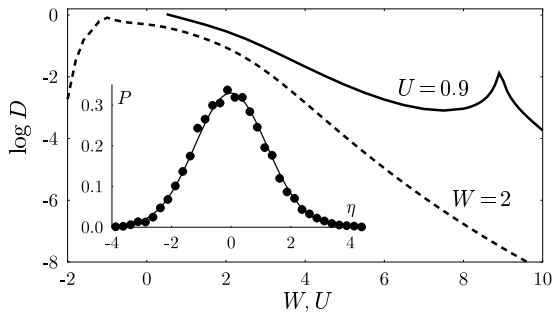}}
\vspace{3mm}
\caption[fig3]{\label{D(W,U)} Phase sensitivity for the sample d of
Fig.~\ref{D(U)}. Solid: $D(W)$ at $U\!=\!0.9$; dashed: $D(U)$ at $W\!=\!2$.
Inset: Probability distribution (dots) of 
$\eta\!=\!\log D(U\!=\!2)\!-\!\log D(U\!=\!0)$ calculated from 10000 samples
($M\!=\!20, N\!=\!10, W\!=\!9$) fitted by a Gaussian with variance 
$\sigma^2\!=\!1.46$.}
\end{figure}
\noindent
the more inhomogeneous density. Repulsive interactions favor the more 
homogeneous density. 
Free fermions correspond to an intermediate case. For small repulsive 
interactions, the system is an Anderson insulator slightly delocalized 
by $U$, and $D(U)$ increases as a function of $U$. At $U_c$, the regular 
array of charges is established, and thereafter it becomes more and more
rigid (pinned by the random lattice); thus $D(U)$ decreases as a function 
of $U$. Certain samples go from the inhomogeneous density to 
the periodic array in a few steps signaled by additional peaks of 
sensitivity. Examining the $U$ dependence of the density 
of those samples, one can note local defects in the periodic array 
subsisting up to large values of $U$. The thresholds $U_c$ are
strongly sample dependent giving rise to a very wide distribution
of phase sensitivities: the ensemble average at a given $U$ mixes very 
different behaviors and provides rather misleading information. As
shown in Fig.~\ref{D(U)}, $\langle\log D(U)\rangle$ decreases for repulsive
interactions, except for a small interval around $U\!\approx\!t$ (inset)
where a local maximum is obtained. Using 5000 samples we confirm beyond 
statistical uncertainty that repulsive interactions delocalize in certain 
parameter region. However, this small average effect is not representative 
of the dramatic enhancement characterizing individual samples.

 We obtain approximately log-normal distributions for $D(U)$ as well
as for the parameter $\eta\!=\!\log D(U\!=\!2)\!-\!\log D(U\!=\!0)$ that 
measures the relative increase of the charge sensitivity with respect
to the free fermion case. The width of the $\eta$-distribution depends
on $U$, and for $U\!=\!2$ we can see that variations of $D$ over
more than an order of magnitude are typical (inset of Fig.~\ref{D(W,U)}).

  For weak disorder ($W\!=\!2, L_1\!\approx\!M$) we recover the expected 
behavior starting from the clean limit, using bosonization and 
renormalization group arguments~\cite{peter,gs2}: a repulsive interaction 
reinforces localization, in contrast to a (not too strong) attractive 
interaction which delocalizes (Fig.~\ref{D(W,U)}). Fixing $U\!=\!0.9$, we 
show (in the same graph) how one goes from the weak to the 

\begin{figure}[tbh]
\centerline{\epsfxsize=3in\epsffile{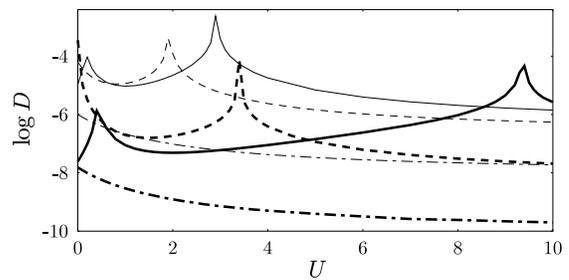}}
\vspace{3mm}
\caption{Phase sensitivity for $M\!=\!40$ and $N\!=\!10$ for 3
samples with $W\!=\!4$ (thin lines) and $W\!=\!5$ (thick lines).}
\label{quarterf}
\end{figure}
\noindent
strong disorder 
limit in a given sample (the same seed of the random number generator is
kept). The phase sensitivity decreases when we increase the strength of 
the potential fluctuations, except around $W\!=\!9$ ($k_{\rm f}l\!\approx\!1$) 
where the charge reorganization takes place.
The conclusion that a repulsive interaction 
favors localization  is no longer valid. We underline that we have 
considered in this study larger disorder than in Ref.~\cite{peter}.
The effect is reminiscent of the Coulomb blockade phenomenon. When the
occupation numbers are approximately good quantum numbers (0 or 1) transport
is blocked; in the transition between such extreme configurations the 
occupation numbers are no longer good quantum numbers and transport is
favored.

 Now, we study if a repulsive interaction can delocalize at a weaker 
disorder for a lower filling. Of course, 
if we keep NN interactions, $U$ will no longer yield 
a fully periodic array (in contrast to a longer range interaction): the 
ground state can partly match the random potential 
without having NN electrons. One expects a more local 
and partial ground state reorganization than the global one discussed 
for half filling. In Fig.~\ref{quarterf}, we show some typical behavior 
obtained at quarter filling for $W\!=\!4$ and $5$. For certain samples 
(dash-dotted), the $U\!=\!0$ Fermi glass does not have NN occupied, 
the charge configuration does not change when turning on the interactions, 
and $D(U)$ decreases with $U$. For others we can observe local 
charge reorganization 
accompanied by an increase of $D(U)$ at one (dashed) or two (solid) values
of $U$. Higher disorder within the same random
configuration (thicker lines) necessitates a stronger interaction $U_c$ to 
produce the charge reorganization. In our model with
NN interactions and high disorder, the increase of the phase 
sensitivity is related to the occurrence of NN in the Fermi glass. Elementary
combinatorics dictates that for a given filling factor $x=N/M$ and large
$M$, the probability of obtaining a configuration without NN is 
$$
P \simeq e^{M g(x)} \ , \hspace{0.5cm} g(x)=\ln{\left(\frac{(1-x)^{2(1-x)}}
{(1-2x)^{(1-2x)}}\right)}\, .
$$
Since $g(x)\!<\!0$ in the interval $(0,1/2)$ of interest, configurations
with NN occur with probability 1 in the large $M$ limit.

\begin{figure}[tbh]
\centerline{\epsfxsize=3in\epsffile{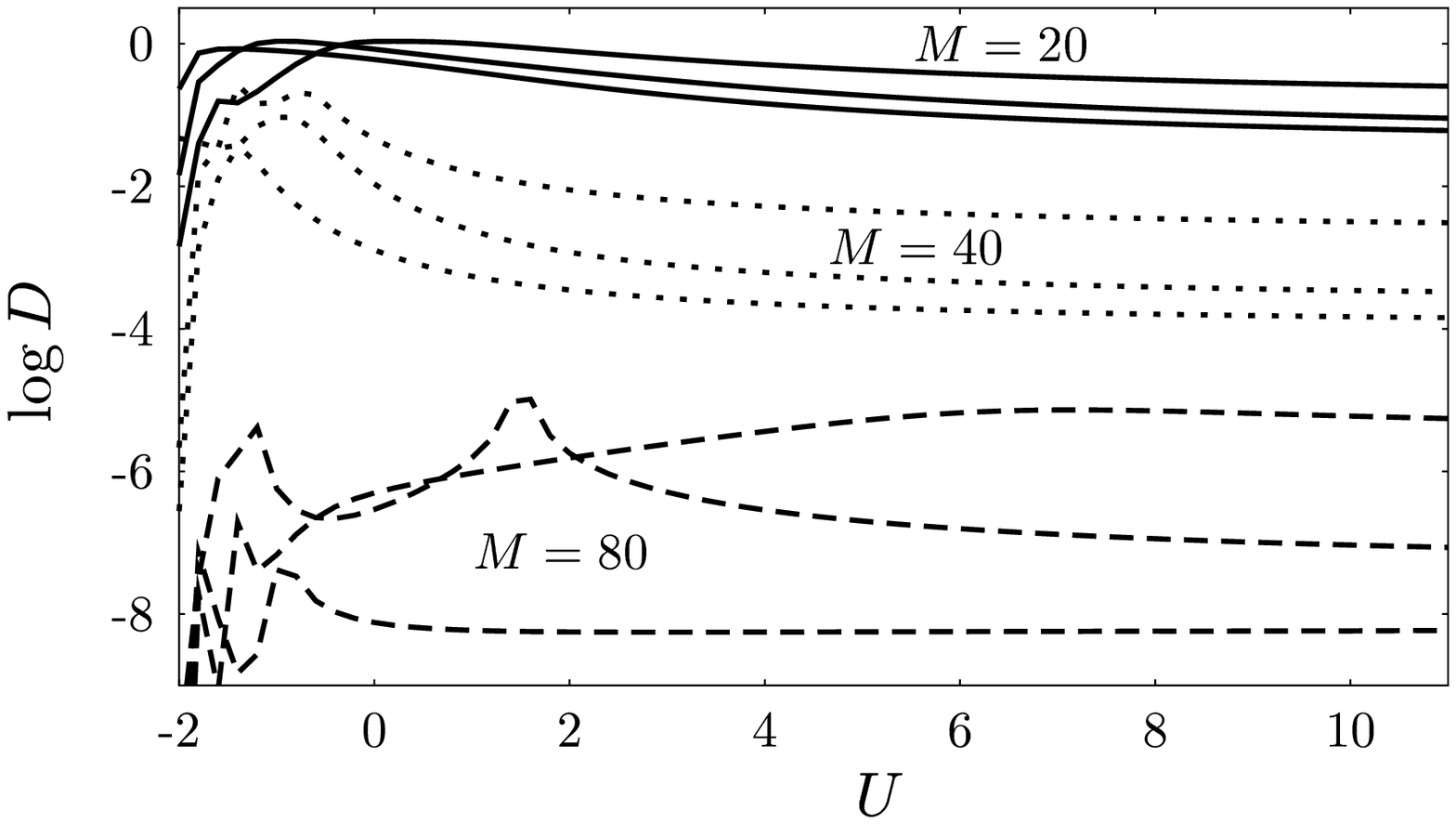}}
\centerline{\epsfxsize=3in\epsffile{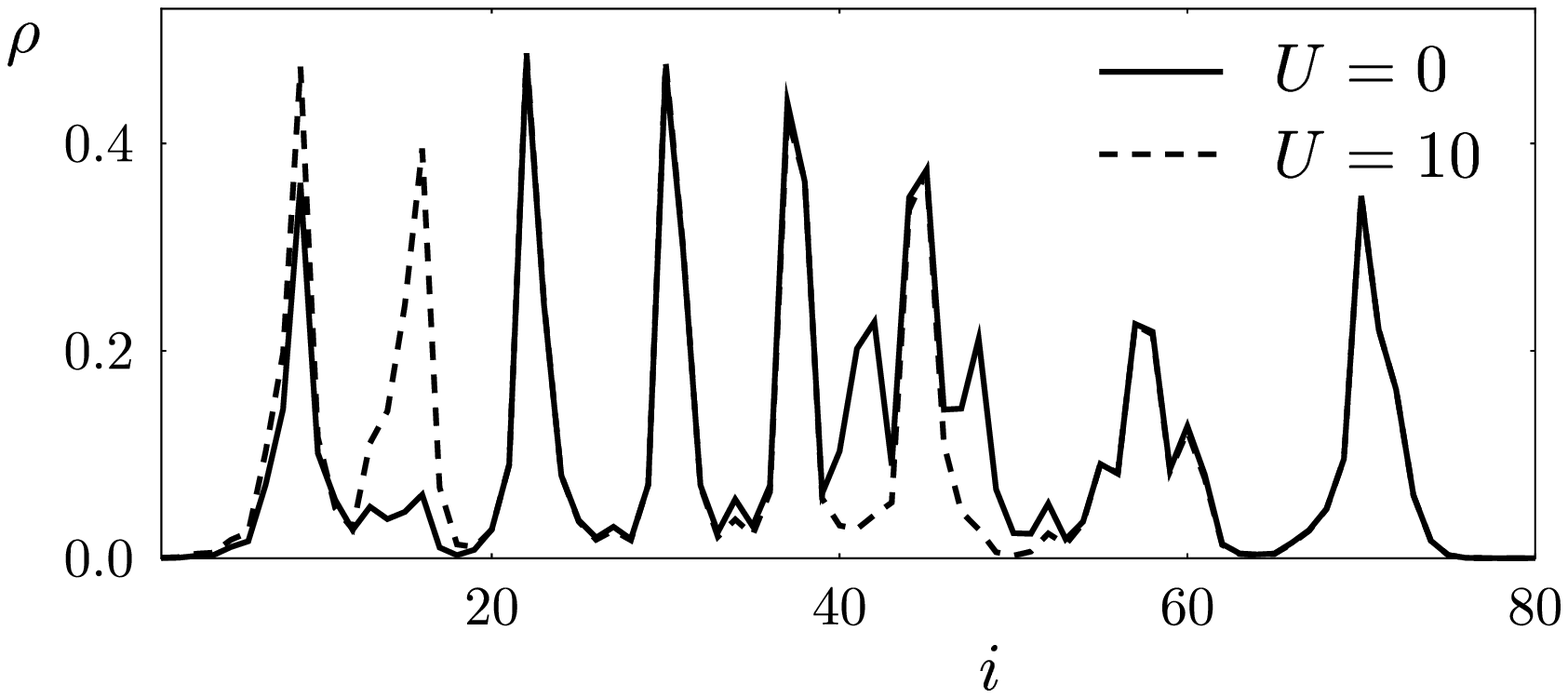}}
\vspace{3mm}
\caption[fig4]{\label{D(L)} Top: phase sensitivity of three samples
with $W\!=\!2$, $N\!=\!8$ and various fillings:
$M\!=\!20$ (solid), $M\!=\!40$ (dotted), and $M\!=\!80$ (dashed).
Bottom: charge density at $U=0$ (solid) and $U=10$ (dashed) for the
sample with $M\!=\!80$ that shows a peak around $U\!=\!1.5$.}
\end{figure}
Fig.~\ref{D(L)} presents $D(U,x)$ at a 
smaller disorder ($W\!=\!2$) for filling factors decreasing down to $1/10$ 
($N\!=\!8$ and $M\!=\!20, 40, 80$). We underline that such a study is 
clearly out of reach if one is restricted to standard diagonalization 
algorithms. To estimate a sample dependent length $L_1$, 
one assumes that $D(U\!=\!0) \approx (M/2) \exp(-2M/L_1)$. This gives for 
$M\!=\!20 (40)$ $L_1\!\approx\!17 (10)$, corresponding to 
$k_{\rm f}l\!>\!1$. In this case a repulsive interaction localizes. 
On the other hand, for a larger 
size $M\!=\!80$, one recovers the delocalization effect when 
$D(U\!=\!0)\approx 10^{-6}$, i.e. when $L_1\!\approx\!9\!\approx M/N$ 
($k_{\rm f}l\!\approx\!1$). The peaks can be now broader and the magnitude of 
the effect is weaker (two orders of magnitude). Studying the charge 
density (bottom), we notice a slow local charge 
reorganization induced by a large variation of $U$. 
When $D(U)\approx 10^{-8}$  (one of the samples with $M\!=\!80$), the 
scale $L_1\!\approx\!7$ of those samples is smaller than $M/N$, $k_{\rm f}l$ 
becomes too small and the Fermi glass cannot be reorganized by a 
NN interaction at this low filling. 

 We summarize the main conclusions that we draw from 
this DMRG study of one dimensional spinless fermions. (i) Each sample 
should be individually studied. The (log)-averages over the ensemble 
are not representative. (ii) The ratio $L_1/(M/N)$ (say $k_{\rm f} l$) defines 
different regimes for short-range repulsive interactions. For 
$k_{\rm f}l\!>\!1$, 
the interaction establishes a correlated array of charges inside $L_1$ 
which is pinned by the random lattice. 
The larger is $U$, the more rigid 
is the array, the more efficient is the pinning and the system is strongly 
insulating (Mott). When $k_{\rm f}l\!<\!1$, the particles can be strongly 
localized far away from each other, and a short range interaction does 
not affect a strongly insulating ground state (Fermi glass). Only very excited 
quasi-particles can be delocalized by the interaction. Between those two 
limits, 
$k_{\rm f}l\!\approx\!1$, the ground state can be deeply reorganized by a 
repulsive interaction $U\!\approx\!t$. This reorganization is accompanied 
by a large delocalization effect. Preliminary results \cite{ben} for  
the interaction induced charge reorganization in two dimensions lead 
to similar conclusions. This might be related to the problem of the 
metal-insulator transition~\cite{krav} observed in Si-Mosfets at very 
low fillings. In quasi-one dimension, this might give some insight for 
the persistent current of a disordered multichannel ring. 

We thank G.\ Ingold, J.-F.\ Joanny, I.\ Safi, X.\ Waintal and 
H.\ Weidenm\"uller for useful discussions. Financial support from 
the TMR network `` Phase coherent dynamics of hybrid nanostructures'' 
of the EU is gratefully acknowledged.

\end{multicols}

\begin{references}

\bibitem{Moriond} 
``Correlated Fermions and Transport in Mesoscopic 
Systems'', Proc.\ of the XXI$^{st}$ Rencontres de Moriond, 
ed.\ by T.\ Martin et al., Editions Frontieres (1996).
 
\bibitem{Levy} 
L.P.~L\'evy et al., {\it Phys.\ Rev.\ Lett.\ }{\bf 64}, 
2074 (1990); V.~Chandrasekhar et al., {\it Phys.\ Rev.\ Lett.\ }{\bf 67}, 
3578 (1991).

\bibitem{nonint}  
A.~Schmid, {\it Phys.\ Rev.\ Lett.\ }{\bf 66}, 80 (1991);
F.~von Oppen and E.~K.~Riedel, {\it ibid} 84; B.~L.~Altshuler, Y.~Gefen,
and Y.~Imry, {\it ibid} 88.

\bibitem{eckern} 
U.~Eckern, {\it Z.~Phys.\ }{\bf B 42}, 389 (1991).

\bibitem{gs} 
T.~Giamarchi and B.~Shastry, {\it Phys.\ Rev.\ }{\bf B 51}, 
10915 (1995).

\bibitem{AetW} 
A.\ M\"uller-Groeling, H.A.\ Weidenm\"uller and C.H.\ Lewenkopf, 
{\it Europhys.\ Lett.\ } {\bf 22}, 193 (1993).

\bibitem{AetB} 
M.~Abraham and R.~Berkovits, {\it Phys.\ Rev.\ Lett.\ }{\bf 70}, 
1509 (1993).

\bibitem{BPM} 
G.~Bouzerar, D.~Poilblanc, and G.~Montambaux, 
{\it Phys.\ Rev.\ }{\bf B 49}, 8258 (1994).

\bibitem{shepelyansky} 
D.L.~Shepelyansky, {\it Phys.\ Rev.\ Lett.\ }{\bf 73} 2607 (1994).

\bibitem{Imry}
Y.~Imry, {\it Europhys.\ Lett.\ }{\bf 30} 405 (1995).

\bibitem{oppen}
F.~von~Oppen and T.~Wettig, {\it Europhys.\ Lett.\ }{\bf 32}, 741 (1995).
 
\bibitem{js}
Ph.~Jacquod and D.L.~Shepelyansky, {\it Phys.\ Rev.\ Lett.\ }{\bf 78}, 4986 
(1997).

\bibitem{wwp}
X.~Waintal, D.~Weinmann and J.-L.~Pichard, cond-mat/9801134. 

\bibitem{gs2}
T.~Giamarchi and H.~Schulz, {\it Phys.\ Rev.\ }{\bf B 37}, 325 (1988). 
 
\bibitem{dmrg}
S.R.~White, {\it Phys.\ Rev.\ }{\bf B 48}, 10345 (1993).

\bibitem{peter}
P.~Schmitteckert et al.,
{\it Phys.\ Rev.\ Lett.\ }{\bf 80}, 560 (1998).

\bibitem{ben}
G.~Benenti, X.~Waintal and J.-L.~Pichard, unpublished. 

\bibitem{krav} 
S.V.~Kravchenko et al., {\it Phys.\ Rev.\ Lett.\ }{\bf 77} 4938 (1996). 


\end{references}
\end{document}